\documentclass[aps,prl,showpacs,twocolumn,floatfix,nofootinbib]{revtex4}
\usepackage{epsfig,graphics,color}
\usepackage{graphicx}
\usepackage{dcolumn}
\usepackage{bm}
\usepackage{amsmath,eufrak}
\usepackage{mathrsfs}

\newcommand \kd  {\delta}

\newcommand \ro {\rho}

\newcommand \e {\epsilon}

\newcommand \x {\cdot}

\newcommand \A {\alpha}

\newcommand \lc {\langle}
\newcommand \rc {\rangle}
\newcommand \prt {\partial}

\newcommand \msf {\mathscr{F}}

\newcommand \sa {\mathscr{A}}
\newcommand \bvec{\left( \begin{array}{c} }
\newcommand \evec{\end{array} \right)}

\newcommand \bea{\begin{eqnarray} }
\newcommand \eea{\end{eqnarray} } 
\newcommand \nn {\nonumber}
\newcommand {\be} {\begin{equation}}
\newcommand {\ee} {\end{equation}}

\newcommand {\mbx} {\mbox{}}

\begin{document}

\title{Possibility of a superfluid in high multiplicity $p$-$p$ and $p$-$Pb$ collisions}

\author{Abhijit Majumder}
\affiliation{Department of Physics and Astronomy, Wayne State University, Detroit, MI 48201.}

\date{\today}

\begin{abstract} 
We consider the case where the saturated and un-equilibrated initial stage of a high multiplicity $p$-$p$ and $p$-$Pb$ collision, 
may, due to high occupation number, display superfluid properties. The special case of a plasma of 
$SU(2)$ gluons formed in the collision of two nuclei with deeply saturated gluon wave functions is considered.
Color gauge symmetry is broken by the presence of a net non-vanishing, space-time dependent color expectation value.
A two component scenario is envisioned, containing momentum states that are densely occupied below the saturation scale 
and a dilute normal fluid of excitations within 
this state. The spectrum of these excitations is shown to have an energy gap and thus it is hard to excite quanta out of the condensed state.
Consequences for the stress energy tensor are outlined, in particular the possibility of near inviscid flow is discussed. 
\end{abstract}

\maketitle

In the last few years, the LHC experiments have been able to carry out systematic studies of high multiplicity $p$-$p$~\cite{Velicanu:2011dp,Abelev:2013bla} 
and $p$-$Pb$ collisions~\cite{Chatrchyan2013795,ABELEV:2013wsa,Abelev:2012ola,Abelev:2013haa}. 
By high multiplicity, we specifically mean that the total number of detected particles is 
$N \gtrsim 100$: a terminology instituted by the experiments. Such events demonstrate long range correlations in rapidity, and have transverse 
momentum spectra and harmonic flow coefficients that can be approximately described assuming that the produced ``medium'' 
obeys almost inviscid fluid dynamics~\cite{Avsar:2011fz,Bozek:2011if,Bozek:2013uha,Shuryak:2013ke,Qin:2013bha}, 
with a viscosity to entropy density ($\eta/s$) similar to that used for heavy-ion collisions.  
There has also been an attempt based on the parton cascade~\cite{Deng:2011at}.  All these efforts are based on the presence of 
a considerable amount of collective flow within the small medium represented by $p$-$p$ and $p$-$Pb$ collisions, which are assumed 
to thermalize after a short time $\tau\!\sim\! 0.6$fm/c. With the exception 
of the parton cascade approach~\cite{Deng:2011at} and the conjecture of transversely expanded initial nucleonic states in 
Ref.~\cite{Qin:2013bha} there has been no attempt to understand the origin of this short thermalization followed by the 
inviscid fluid behavior in these systems.

In this Letter, a microscopic description of the medium produced in such systems is proposed. While some manifestation of this 
picture may also be applicable to heavy-ion collisions, we refrain from making such a connection at this point. The mechanism 
proposed in this letter should, currently, be applied solely to $p$-$p$  and $p$-$Pb$ collisions (we are specifically referring to 
collisions at LHC energies where very high multiplicity events have been detected). 
Our central premise is that the initial state in very high multiplicity $p$-$p$ and $p$-$Pb$ collisions has the properties of a 
superfluid, due to very high occupation number of gluons in all states up to a scale of the order of the saturation scale $Q_{S}$, and a dispersion 
relation of excitations that makes it difficult to excite gluons out of this state. In what follows, the causes for the appearance of
this state will be outlined. An effective Lagrangian for the simple case of quark-less $SU(2)$ containing the condensate of 
gluons and small amplitude excitations will be surmised. The dispersion relation for the excitation spectrum will be derived. 
Consequences for the ensuing form of the stress-energy tensor ($T^{\mu \nu}$) will be outlined.

It is now widely accepted that the initial state of $p$-$p$, $p$-$A$ collisions (and also $A$-$A$, with $A$ representing a large nucleus) 
contains a very high density of virtual gluons which populate the low forward momentum, or low-$x$, part of the nucleon or 
nuclear wave-function ($x=p^{+}/P^{+}$, where $p^{+} = ( p_{0} + p_{z} )/\sqrt{2}$ is the parton light cone momentum and $P^{+}$ is that for 
a nucleon)~\cite{McLerran:1993ni,McLerran:1993ka,Mueller:1993rr}. We are assuming the colliding beams to be traveling in the $z$ direction.
While the gauge fields in this configuration, called the ``color 
glass  condensate'' are mostly transverse due to the large boost of the nuclei, the fields immediately after the collision have both longitudinal 
and transverse polarizations, carry topological charge, and are commonly referred to as the ``glasma''~\cite{Lappi:2006fp}. 
Fluctuations in the glasma have been related to the long range correlations in the rapidity called the ``ridge''~\cite{Dumitru:2008wn}, recently seen in 
high multiplicity $p$-$p$ and $p$-$Pb$ collisions, and most prominantly in heavy-ion collisions. 

Attempts to match the initial conditions that arise 
from the dynamics of the glasma to the start of hydrodynamic evolution have also been made~\cite{Schenke:2012wb,Chen:2013ksa}.
There has also been the attempt to argue that in the course of thermalization of the glasma, due to the overpopulation of soft modes with 
momentum below the saturation scale $Q_{S}$, a 
gluonic Bose-Einstein condensate (BEC)~\cite{Blaizot:2011xf} would arise in such systems.
The work in this Letter, is somewhat related to the latter work mentioned above, however it neither requires thermalization, nor 
Bose-Einstein condensation to demonstrate the presence of superfluidity in such systems. 
There is also no requirement that full saturation dynamics have set in, only that there is a large enough density of 
softer gluons that one may use classical fields to describe this state.

Imagine the collision of two protons at very high energy. Both wave functions of the two incoming protons possess gluon distributions that are 
saturated up to a transverse momentum of $k_{\perp} \lesssim Q_{S} \sim \alpha_{S} N_{c} x G(x)/(\pi R^{2})  $, where $\A_{s}$ is the strong 
coupling constant, $N_{c}$ is the number of colors, $G(x)$ is the gluon distribution function and $R$ is the transverse radius of the proton. 
All these gluons represent colored fluctuations from the high momentum (hard) color charges inside the proton. Those that populate the mid-rapidity 
or $y = \log \left[ \frac{p^{0} + p_{z}}{p^{0} - p_{z}}\right]/2 = 0$ region at LHC collisions with $\sqrt{s} \sim 5$TeV correspond to an $x \sim 10^{-4}$.
The prevalent state immediately after the collision, called the glasma, is formed by the fusion of such low-$x$ virtual gluons from the two nuclei, resulting in  
near on-shell gluons with $k_{\perp} \lesssim Q_{s} $. At mid-rapidity, these gluons also have a $k_{z} \ll k_{\perp}$, as gluons with larger $k_{z}$ will appear at 
larger rapidities. In this Letter we consider the small rapidity region of these collisions at a time $\tau$ such that $1/\Lambda_{QCD} \! \gg \! \tau \! \gg \! 1/Q_{S}$. 
This of course requires $Q_{S} \gg \Lambda_{QCD}$, which is only true in very high energy collisions. 
  
Without going into further details of the production of the glasma, we consider its properties in very high multiplicity events. 
%
%
As a first simplification, we consider all modes from the 
lowest momentum scale up to a scale $Q \sim Q_{S} $ to be considerably over populated, such that one can apply classical field theory to the description of this condensed state. 
Modes with momenta 
above $Q$ will be considered as a dilute gas which may interact with and produce excitations on the saturated glasma and eventually equilibrate with it.
As such we may write down an effective Lagrangian for the saturated soft modes and excitations of this state by separating the non-abelian vector potential as 
a classical part (soft modes) and a small fluctuation (hard modes),
\bea
A^{a}_{\mu} = A^{a}_{\mu} + \sa^{a}_{\mu}.
\eea
One can now expand the gauge field Lagrangian as a series in the small fluctuation field $\sa^{a}_{\mu}$, as
\bea
\mathscr{L}_{0} &=& -\frac{ F^{\mu \nu} F_{\mu \nu} }{4},  \\
\mathscr{L}_{1} &=& -\frac{ \msf^{\mu \nu}   \msf_{\mu \nu}}{4} + \mathscr{V}. 
\eea
In the equations above $F^{\mu \nu} = t^{a} ( \prt^{\mu} {A^{a}}^{\nu} - \prt^{\nu} {A^{a}}^{\mu}  + g \e^{abc} { A^{b} }^{\mu} { A^{c} }^{\nu}  )$ and 
$\msf^{\mu \nu} = t^{a} ( \prt^{\mu} {\sa^{a}}^{\nu} - \prt^{\nu} {\sa^{a}}^{\mu}   )$. Interaction terms between the fluctuation field and the condensate 
are contained within $\mathscr{V}$,  expressed as, 
\bea
\mathscr{V} \!\!&=&\!\! \frac{-g^{2}}{4} \left[ 2 \e^{abc} { A^{b} }^{\mu}  { A^{c} }^{\nu}  \e^{a d e} \sa^{d}_{\mu} \sa^{e}_{\nu} 
+  \e^{abc} \left(  { A^{b} }^{\mu}  { \sa^{c} }^{\nu}  \right. \right. \nn \\
&+&\!\!  \left. \left. { \sa^{b} }^{\mu} { A^{c} }^{\nu}  \right)  \e^{a d e} \left( A^{d}_{\mu} \sa^{e}_{\nu}  + \sa^{d}_{\mu} A^{e}_{\nu} \right)  \right].
\eea
In the equation above, $\lc \sa^{\mu} \rc = 0$ by definition and we have dropped cubic and quartic terms in the fluctuation field. 
We have also dropped all 
terms that involve derivatives of the condensate. While these are not vanishing, over the range of distances and times inhabited by the fluctuation 
field, they are small compared to the derivatives of 
the fluctuation field (as well as the amplitude of the condensate wave-function ${ A^{a} }^{\mu}$). In the writing of the above set of equations, we have 
also simplified the gauge group to $SU(2)$.  This is done primarily to simplify the ensuing derivation. With the exception of a one-dimensional weight 
diagram compared to a two-dimensional weight diagram, no major differences are expected between this and the more physical case of $SU(3)$.

We now introduce approximations regarding the condensate within the effective theory. 
Rewriting the interaction term between the classical field and the hard fluctuation, we obtain, 
\bea
\mathscr{V} = - \frac{g^{2}}{2} \sa^{b}_{\mu}  \mathcal{S}^{(bc) \mu \nu } \sa^{c}_{\nu}, \label{V-and-S}
\eea
where, all factors of the classical field are contained within $\mathcal{S}^{(bc) \mu \nu } $.
In the effort to describe the tree level behavior of the fluctuation field in the presence of an average soft background field, an event 
average over the product of background fields in $ \mathcal{S}^{(bc) \mu \nu } $ is carried out. As a result, we use the following 
simple approximation for the product of gauge fields, 
\bea
\langle A_{\mu}^{a} A_{\nu}^{b} \rangle \simeq \frac{\kd^{ab}}{N_{c^{2}} - 1}  \frac{g_{\mu \nu}}{4}  \langle A_{\A}^{a} A^{a\A } \rangle .
\eea
In this Letter, the interaction of the fields will be described using axial gauge $A^{a}_{3} = \sa^{a}_{3} = 0$ . In this gauge, 
the $A^{a}_{0} = \phi^{a}$ field is derived from the equations 
of motion. The fluctuation $\sa^{a}_{0}=\psi^{a}$ is introduced as a Lagrange multiplier to integrate out the conjugate momenta 
using completion of squares. 

To determine the effect of the interaction term on the fluctuation field, we solve the classical equations of motion for the background 
field,
\bea
\mathcal{D}_{\mu}^{ac} F^{c \mu \nu} = \left( \kd^{ac} \prt_{\mu} + g \e^{abc} A^{b}_{\mu} \right) { F^{c} }^{\mu \nu} = 0. 
\eea
Over the times and distances traversed by the higher frequency fluctuation fields, one may ignore all derivatives of the classical 
background field in comparison to the amplitude of the fields themselves. As a result, the equations of motion simplify: e.g. for the case of 
$\nu =0 $ to,
\bea
\e^{abc} A^{b}_{\mu} \e^{cde} A^{d \mu} \phi^{e 0} \simeq 0. \label{EOM-approx-1}
\eea
For $\nu = 3$, due to the choice of axial gauge, the leading term has one derivative, 
\bea
\e^{abc} A_{\mu}^{b} \prt_{3} A^{c \mu} \simeq 0. \label{EOM-approx-2}
\eea
The other components $\nu = 1,2$ will yield equations similar to Eq.~\eqref{EOM-approx-1}. 

Since the glasma gauge fields have been sourced by the virtual gauge fields in the colliding nucleons, the initial conditions are determined in 
every event given a model of the transverse gluon fields in the colliding nucleons. 
The equations above, describe the behavior of the scalar potential $\phi^{a}$ and the behavior of the transverse gauge fields over short times after the collision. 
Over longer time periods and larger distances, the higher derivatives will no longer be negligible.

Our goal here is to demonstrate the possibility of 
of a mass gap for excitations with momenta above the saturation scale. We now introduce an \emph{ansatz} for the classical 
field which will include a condensate. Here we follow the work of Refs.~\cite{Alford:2012vn,Sannino:2002wp}, and decompose the 
classical vector field as
\bea
{A^{+}}^{j} &=& \frac{ {A^{1}}^{j} + i {A^{2}}^{j} }{\sqrt{2}} = \rho^{j} e^{i \zeta}  \,\,\,\,\,\,\,\,\, (j = 1,2), \\
{A^{-}}^{j} &=& \frac{ {A^{1}}^{j} - i {A^{2}}^{j} }{\sqrt{2}} = \rho^{j} e^{-i \zeta} \,\,\, \& \,\,\, {A^{3}}^{j} = Z^{j},
\eea
where, $\rho = \sqrt{  \rho_{x}^{2} + \rho_{y}^{2} }$ is the amplitude of the condensate and $\zeta$ is the phase (We have tacitly assumed 
the same phase for both the $x$ and $y$ components: $\rho_{x} = \rho \cos{\theta}$  and $\rho_{y} = \rho \sin{\theta}$). 
The projection of $\ro^{j}e^{\pm \zeta}$ and $Z^{j}$ along the ($x,y$)-axis are assumed to factorize, i.e., $\ro^{j} = \rho \hat{a}^{j}$ and 
$Z^{j} = Z \hat{b}^{j}$, where $\hat{a}$ and $\hat{b}$ are 2-dimensional unit vectors.
The presence of a condensate naturally breaks both color symmetry and 
rotational symmetry. In this first attempt we will ignore any effect of the breaking of rotational symmetry. The symmetries are 
broken explicitly (by the collision of two nucleons) and not spontaneously. As a result there are no massless Goldstone bosons. 

One can also decompose the scalar potential in a fashion similar to the transverse fields, 
\bea
\phi^{+} = Ve^{i\eta} \,\, , \,\, \phi^{-} = Ve^{-i \eta} \,\, , \,\, \phi^{3} = U,
\eea
i.e., the scalar potential can be out of phase with the transverse gauge fields. Substituting these back into 
Eqs.~(\ref{EOM-approx-1},\ref{EOM-approx-2}) along with the equation for $\nu=3$ and obtain two 
sets of solutions. One of these solutions leads to negative total energy and this will be ignored. In what follows 
we focus on the positive energy solution. 
We obtain equations for the gauge fields and their derivatives from Eqs.~(\ref{EOM-approx-1},\ref{EOM-approx-2}), by equating real and imaginary parts; these yield
$V = \rho$, and 
\bea
U = \frac{ 2 \rho^{2} Z \cos \left( \zeta - \eta \right) \hat{a}\x\hat{b} } {2 \rho^{2} + Z^{2} (1 - \hat{a} \x \hat{b}) }.
\eea
The two phases are related to each other as $\eta = \zeta + \Delta \phi$, where $\prt_{3}\Delta \phi =0$, and $\Delta \phi$ is 
small enough that $\sin{\Delta \phi} \sim \Delta \phi$.
These approximations, along with the event average taken above, reduce the factor $\mathcal{S}^{(bc)}_{ \mu \nu }$ in Eq.~\eqref{V-and-S} to its diagonal form $g_{\mu \nu} \kd^{bc} \mathcal{S}/[4 ( N_{C}^{2} - 1 ) ]$, with $\mathcal{S}$ given as, 
\bea
\mathcal{S} = - Z^{2}  \left\langle 1 - \frac{ 4 \rho^{4} \cos^{2} \left( \zeta - \eta \right) (\hat{a} \cdot \hat{b} )^{2}  } { \left[ 2 \rho^{2} + Z^{2} (1-\hat{a} \cdot \hat{b})  \right]^{2} } \right\rangle
\eea
As a result, on average $\mathcal{S} < 0$ and this term behaves in the fluctuation Lagrangian as an $m^{2}$ term (i.e., a mass term at tree level).
As a result, tree level dispersion relations of the fluctuation field $\sa^{a \mu}$, with the inclusion of an interaction with the mean condensate will posses 
an energy $E  > p$, the momentum of the modes. Hence such modes will be difficult to excite and thus the condensate will be unable to equilibrate by
populating the higher momentum modes. 
The incorporation of this mass term in the propagators of the 
fluctuation field is akin to the Bogoliubov transformation in condensed Bose systems~\cite{bogolubov1947theory} and similar to dispersion relations required in the recent condensation of photons~\cite{RevModPhys.85.299}. While the analysis in this 
Letter is carried out in axial gauge, we expect a mass correction to arise in other gauges as well. Such a correction is seen to arise also in the simplified charged scalar theory~\cite{Kapusta:1981aa}.

As a result of this condition, at small rapidities in $p$-$p$ and $p$-$A$ collisions, for a time $\tau > 1/Q_{S}$, there may exist a superfluid phase in the glasma (at larger 
rapidities this phase may appear at a later time). 
The superfluid four-velocity is given by the 
derivative of the phase of the condensate~~\cite{Alford:2012vn},
\bea
u^{\mu} = \frac{ \prt^{\mu} \zeta }{\sqrt{ \sigma}}  = \frac{ \prt^{\mu} \zeta }{\sqrt{ \prt^{\nu} \zeta \prt_{\nu} \zeta }} .
\eea
As the glasma expands, the magnitudes of the condensates $Z,\rho$ will tend to drop and the magnitude of the energy gap will drop. As a result, with increasing time, it will become progressively easier to excite modes out of the condensate leading to a diminishing of the condensate and population of the ``normal fluid'' of excitations (hard modes).  
For small systems produced in $p$-$p$ and $p$-$A$ collisions, it is not clear if there will be a further inviscid fluid phase due to the strong 
interactions in the normal fluid. It is also interesting to note that with a drop in the amplitude of the gauge fields, the separation scale $Q \sim Q_{S}$ will also 
drop with time and thus the fluid will continue to become more strongly interacting with increasing time. If the superfluid were to persist past equilibration, this mechanism would provide another reason for 
the appearance of a BEC as described in Ref.~\cite{Blaizot:2011xf}.
To determine the fate of the superfluid phase and the possible appearance of a normal fluid phase would require a numerical simulation which is beyond the scope of this Letter.

In the limit that higher derivatives of the condensate can be ignored, in particular, second derivatives, and if the condensate fraction dominates over the normal fluid, the energy-momentum tensor has a simple form in terms of only the gauge fields. Here we list a few of the components obtained from the 
expression for the traceless part of the classical energy momentum tensor of a non-abelian theory (we ignore issues related with 
operator renormalization):
\bea
\mbx\!\!\!\!{\Pi}^{00} &=& g^{2} ( \Delta \phi)^{2} \left[  (2 \rho^{2} + Z^{2} )\rho^{2} \right], \nn \\
\mbx\!\!\!\!\Pi^{ii} &=& g^{2} [2 \rho^{2} + Z^{2}](\Delta\phi)^{2} \left[  \rho^{2} (1 - 2\hat{a}^{i}\hat{a}^{i})  \right], \nn \\
\mbx\!\!\!\!\Pi^{33} &=& = g^{2} [2 \rho^{2} + Z^{2}](\Delta\phi)^{2}  \left[  \rho^{2} \right], \nn \\
\mbx\!\!\!\!\Pi^{0i} &=& g^{2} \hat{a}^{i} [2 \rho^{2} + Z^{2}](\Delta\phi)^{2} \rho^{2} \,\, , \,\, \Pi^{03} = 0 \,\, {\rm etc.}
\eea
Such an expression for the stress-energy tensor is only valid in the limit that the amplitude of the gauge fields is very large compared to both the 
contribution from the normal fluid and from derivative terms in the condensate. With the inclusion of the first set of derivative terms, one will 
obtain both derivatives of the amplitude of the condensate $(\rho, Z)$, as well as derivatives of the phase $\phi$, which will yield factors of 
velocity $u^{\mu}$. Even at this stage, the viscosity of the fluid will be effectively vanishing. With the rise in the population of the excited 
modes(normal fluid), as well as the magnitude of second derivatives of the condensate, viscous terms will appear in the stress energy tensor of this field. As a result, we may hypothesize that very high multiplicity $p$-$p$ and $p$-$A$ collisions will behave like an inviscid fluid from a time 
$\tau > 1/Q_{S}$, and continue to retain this behavior until there is a sufficient population of the excited states. The determination of the exact 
time when this will take place requires a numerical simulation of the system, which we leave for a future effort.

In this Letter, we have outlined a possible reason for the appearance of inviscid fluid behavior in high multiplicity $p$-$p$ and $p$-$A$ collisions: 
The appearance of a colored superfluid of gluons. The analysis carried out in this letter is very simplified, the goal was to highlight the means by which 
an energy gap in the excitations of the normal fluid may arise by interaction with the condensate. The persistence of such large fields leads to vanishingly small viscosity in such systems and may be the underlying reason for the perfect fluid nature in such systems. While such a superfluid would appear a very short time after the collision, 
the time up to which it will persist is uncertain at this point and will require a more detailed numerical analysis. 

Discussions with S. Gavin, U. Heinz, B. Nadgorny, G. Paz, A. Petrov, and J. Putschke are gratefully acknowledged.
This work was supported in part by the NSF under grant number PHY-1207918.

\bibliography{refs}

\end{document}